
\def\sqr#1#2{{\vcenter{\hrule height.#2pt
             \hbox{\vrule width.#2pt height#1pt \kern#1pt
             \vrule width.#2pt}
             \hrule height.#2pt}}}
\def\boxx{\hskip1pt\mathchoice\sqr54\sqr54\sqr{3.2}4\sqr{2.6}4
                                               \hskip1pt}
\font\eightrm=cmr8               
\font\bfi=cmmib10                
\font\eightbf=cmbx8
\def\pmb#1{\setbox0=\hbox{#1}\kern-.025em
    \copy0\kern-\wd0\kern.05em\kern-.025em\raise.029em\box0}
     
     \def\mb#1{\hbox{\bfi #1}}
\def\sc#1{\hbox{\sc #1}}

\def\a{\alpha}      \def\l{\lambda}   
\def\b{\beta}       \def\m{\mu}
\def\g{\gamma}      \def\G{\Gamma}    \def\n{\nu}
\def\d{\delta}      \def\D{\Delta}         
\def\e{\epsilon}                      
\def\ve{\varepsilon}                  \def\s{\sigma}  \def\S{\Sigma}

                   \def\o{\omega}   
\def\k{\kappa}

\def\fr#1 #2{\hbox{${#1\over #2}$}}
\def\section#1{
  \vskip.7cm\goodbreak
  \centerline{\bf \uppercase{#1}}
  \nobreak\vskip.4cm\nobreak  }
\def\subsection#1{
  \vskip.7cm\goodbreak
  \noindent{\bf #1}
  \nobreak\vskip.4cm\nobreak  }
\def\vidi#1{
   {\vskip.3cm \noindent
    \parindent=20pt \narrower
    {\sl #1}
    \vskip.3cm}        }
\def\subsub#1{\par\vskip.3cm {\bf #1} }

\def\ftn{\eightrm\baselineskip=9pt}

\def\aut#1{{\sl #1}}
\def\rad#1{"#1"}
\def\knj#1{{\it #1}}

\magnification=\magstep1
\null

\section{The Liouville theory and $SL(2,R)$ symmetry $^\ast$}
\footnote{}{$^\ast\,$\ftn Lectures presented at the Danube Workshop '93,
June 1993, Belgrade, Yugoslavia.}

\vskip.5cm
\centerline{\it M. Blagojevi\' c}
\centerline{\it Institute of Physics, P.O.Box 57, 11001 Belgrade,
                                                            Yugoslavia}
\vskip.5cm

\centerline{\eightbf Abstract}
\vskip.2cm
{\eightrm The Liouville action emerges as the effective action of 2-d
gravity in the process of path integral quantization of the bosonic
string. It yields a measure of the violation of classical symmetries
of the theory at the quantum level. Certain aspects of the residual
$SL(2,R)$ invariance of the gauge-fixed Liouville theory are
disscussed.}

\subsection{1. Polyakov's path integral and the Liouville action} 

String moves trough a D-dimensional spacetime $X_D$, sweeping out a
two-dimensional surface $\S$. In the Polyakov version, the classical
action of the bosonic string takes the form
$$
I[x^\m ,g_{\a\b}]=\int d^2\xi\sqrt{g}g^{\a\b}\partial_\a x^\m
                  \partial_\b x^\n G_{\m\n} \, .           \eqno(1)
$$
The worldsheet metric $g_{\a\b}$ as well as $x^\m$ are treated as {\it
independent dynamical variables}, while the metric of spacetime
$G_{\m\n}$ is {\it fixed} (we shall assume that
$G_{\m\n}=\eta_{\m\n}$).  The action (1) is classically equivalent to
the Nambu action.  The equivalence of the two formulations at the
quantum level is a more comlicated question. The Polyakov action is
more convenient for the covariant path integral quantization.

In order to be able to apply the mathematical theory of Riemann
surfaces we shall work with the Euclidean worldsheet, which is obtained
by the Wick rotation from the original Minkowskian worldsheet. After
all the calculations are completed, one may return to the Minkowskian
theory by analytic continuation.

\subsub{The functional integral.}  We shall limit our discussion to the
usual {\it perturbative approach}. Thus, a scattering process for $n$
strings in the approximation of $l$ loops is described by a
two-dimensional manifold (surface) of {\it genus} $\g=l$ and with $b=n$
{\it boundary components}.  The scattering amlitude is assummed to be
of the form [1-3]
$$
A_\g (p_1,p_2,\dots ,p_n)=\int_{\S_{\g ,n}}Dg_{\a\b}Dx^\m e^{-I[x,g]}\,
                      V(p_1)V(p_2)\dots V(p_n)\, ,         \eqno(2)
$$
where $V(p_i)$ are vertex operators assigned to the boundaries of $\S$,
and $\int Dx^\m$ is the integration over
all continuous maps $x^\m (\xi )$ from $\S$ to $X_D$. The complete
amplitude for a given process is obtained by summing over all $\g$'s in
Eq.(2).

The amplitude with $n=0$ (no external string states), decribing the
{\it vacuum-to-vacuum transition},
$$
Z=\sum_\g \,\int_{\S_\g}Dg_{\a\b}Dx^\m e^{-I[x,g]} \, ,    \eqno(3a)
$$
will be the main subject of our discussion.

We shall concentrate on the {\it closed, oriented} bosonic string
theory.  We shall also restrict the surfaces in (3a) to be {\it
compact, closed} and {\it oriented} (and connected, as we are
interested in the connected part of the amplitude). Finally, we shall
discuss mainly the contribution of {\it genus zero} surfaces, whose
description is technically the simplest:
$$
Z_0=\int_{\S_0}Dg_{\a\b}Dx^\m e^{-I[x,g]} \, .             \eqno(3b)
$$
Several mathematical theorems related to the geometry of
two-dimensional surfaces are given in Appendix A.

\subsub{The measure $Dg_{\a\b}$}. The classical string action is
invariant under Weyl rescalings and diffeomorphisms. As a consequence,
the functional integral $Z_0$ is highly divergent. The elimination
of this divergence demands a carefull treatment of the measure,
as in the case of gauge theories (the Faddeev-Popov construction).

The integration over all metrics in Eq.(3b) has the following
meaning. All genus zero manifolds $\S_0$ are topologically
equivalent to each other, but different as metric spaces
$(\S_0,\mb{g})$. With the topology of $\S_0$ fixed, the set of all
metric spaces $(\S_0,\mb{g})$ is determined by the set of all
addmissible metrics on $\S_0$:
$$
{\cal M}_0(\mb{g})=\{ \mb{g}\,\vert\,\mb{g}
                                    \hbox{ is a metric on }\S_0\}\, .
$$
The measure $Dg_{\a\b}$ can be determined by introducing the metric
\mb{G} on ${\cal M}_0(\mb{g})$.

{\it Example}: In an n-dimensional Riemannian space with metric
$ds^2=G_{ab}dy^ady^b$, the masure is defined as
$\sqrt{G}dy^1\dots dy^n$.

Let $\d\mb{g}$ be a tangent vector at a point $\mb{g}\in
{\cal M}_0(\mb{g})$ with components $\d g_{\a\b}(\xi )$. The scalar
product
$$
(\d\mb{g}^{(1)}\, ,\d\mb{g}^{(2)})=\int d^2\xi\sqrt{g}g^{\a\g}g^{\b\d}
                 \d g^{(1)}_{\a\b}\d g^{(2)}_{\g\d}        \eqno(4)
$$
defines a natural metric in the tangent space ${\cal T}_0(\mb{g})$.
This metric is invariant under reparametrizations, but {\it not under
Weyl rescalings\/.} It defines the measure $Dg_{\a\b}$.

All genus zero metric spaces $(\S_0,\mb{g})$ are {\it conformally
equivalent}, i.e. any two metrics are related to each other by the
local diffeomorphisms and Weyl rescalings only. This follows from
{\bf T4} of Appendix A, and holds only for $\g =0$ (if $\g\ge 1$ one
has also global diffeomorphisms, and the descrete groups $\G_\g$ are
nontrivial). Therefore,
\vidi{choosing a particular metric
$\hat{\mb{g}}$ in ${\cal M}_0(\mb{g})$, an arbitrary metric
\mb{g} can be obtained from $\hat{\mb{g}}$ by Weyl
rescalings and local diffeomorphisms only.}
To be specific we can choose $\hat{\mb{g}}$ to be conformally flat,
i.e.
$$
{\hat g}_{\a\b} = e^{\hat\phi}\d_{\a\b} \, .
$$
This choice can always be made, as we know from {\bf T2}
(there are no essential changes in the discussion which follows if
$\hat{\mb{g}}$ is chosen differently). The set of metrics that are
connected to $\hat{\mb{g}}$ by Weil rescalings defines the Weyl slice
in ${\cal M}_0(\mb{g})$; if $\hat{\mb{g}}$ is conformally flat, the
corresponding Weyl slice is called the conformally flat gauge slice.
It is easy to see that Weyl transformations
$$
\d g_{\a\b} = \d\phi\,g_{\a\b}                             \eqno(5)
$$
slide the conformally flat metric ${\hat g}_{\a\b}$ along the
conformally flat gauge slice.


The local diffeomorphisms $Diff_0$ have the form
$$
\d g_{\a\b}=\nabla_\a \e_\b + \nabla_\b \e_\a \, .
$$
For $\g =0$ all diffeomorphisms are connected to the identity.
These transformations  can be decomposed into the traceless and trace
part
$$\eqalign{
&\d g_{\a\b} = \d^D g_{\a\b} + (\nabla\cdot\e )g_{\a\b} \, ,\cr
&\d^D g_{\a\b}\equiv \nabla_\a \e_\b + \nabla_\b \e_\a
           -g_{\a\b}(\nabla\cdot\e )\equiv (P_1\e )_{\a\b} \, .}
$$
The trace part is tangent to the Weyl slice containg $\hat{\mb{g}}$.
Combining $Diff_0$ with Weyl rescalings one obtains
$$\eqalign{
&\d g_{\a\b} = \d^W g_{\a\b} + \d^D g_{\a\b} \, , \cr
&\d^W g_{\a\b}\equiv (\d\phi +\nabla\cdot\e )g_{\a\b} \equiv
                     \d\l \,\, g_{\a\b} \, .}              \eqno(6)
$$
The tanget vectors $\d^W\mb{g}$ and $\d^D\mb{g}$ are {\it orthogonal}
with respect to the scalar product (4); they {\it span} the tangent
space to ${\cal M}_0(\mb{g})$ at the point $\hat{\mb{g}}$ (no other
transformations of metric are admissible in the $\g =0$ case).

Our intention is to construct the measure $Dg_{\a\b}$ in terms of
parameters $(\d\phi ,\e)$ associated with Weyl rescalings and $Diff_0$,
which take us from $\hat{\mb{ g}}$ to \mb{g}. The norm of the tangent
vector $\d\mb{g}$, defined in Eq.(6), can be calculated by using the
scalar product (4):
$$
\Vert \d\mb{g}\Vert^2 = (\d\mb{g}\, ,\d\mb{g}) = \int d^2\xi\sqrt{g}
       [ (\d\l )^2 +(P_1\e )_{\a\b}(P_1\e )^{\a\b}]\, .
$$
Note that this norm is reparametrization invariant but {\it not Weil
invariant\/,} which will result in the loss of Weil invariance of the
functional integral (quantum theory).  After introducing norms in the
{\it tangent spaces} of scalars and vectors as
$$
\Vert\d\l\Vert^2 \equiv \int d^2\xi\sqrt{g}(\d\l )^2 \, ,\qquad
\Vert\d v\Vert^2 \equiv  \int d^2\xi\sqrt{g}g_{\a\b}\d v^\a\d v^\b \, ,
$$
we can write
$$
\Vert \d\mb{g}\Vert^2 = \Vert \d\l\Vert^2 + \Vert P_1\e \Vert^2\, .
$$
Since
$$
\Vert P_1\e \Vert^2 = (P_1\e ,P_1\e) = (\e ,P^+_1 P_1\e ) \, ,
$$
we could conclude that the measure $Dg_{\a\b}$ is given by
$$
Dg_{\a\b}= \sqrt{G}D\phi D\e^\a = [\det (P^+_1 P_1)]^{1/2}
                                          D\phi D\e^\a \, . \eqno(7a)
$$
However, this is not true since $P_1$ has {\it zero modes}. Indeed, the
equation
$$
(P_1\e )\equiv \nabla_\a \e_\b + \nabla_\b \e_\a
           -g_{\a\b}(\nabla\cdot\e ) = 0
$$
is the conformal Killing equation whose nontrivial solutions are the
Killing vectors $\e^K_\a$. The diffeomorphisms generated by $\e^K_\a$
have the same effect as a Weyl transformation. Therefore, the measure
takes the form
$$
Dg_{\a\b}= [\det{'}(P^+_1 P_1)]^{1/2}D\phi D\e^{\prime\a}  \eqno(7b)
$$
where prime means that the zero modes have been {\it deleted}, and the
functional integral (3b) becomes
$$
Z_0= N \int D\phi D\e^{\prime\a} Dx^\m [\det{'}(P^+_1 P_1)]^{1/2}
     e^{-I[x,g]} \, ,                                      \eqno(8)
$$
where $g_{\a\b}=g_{\a\b}(\phi,\e^{\prime\a})$.

We want to extract the volume $V(Diff_0)$, which differs from
$V'(Diff_0)= \int D\e'$ by the contribution of the Killing vectors,
$V^K(Diff_0)$. Since the two types of the diffeomorphisms are mutualy
orthogonal, we have $V=V'\cdot V^K$. The number of independent Killing
vectors in the case $\g =0$ is $6$ (the number of independent
parameters of $SL(2,C)$, the conformal group of the compactified
complex plane $\hat C$). Introducing a basis $\mb{e}_a (a=1,\dots
,6)$,we can write $V^K =[\det (\mb{e}_a,\mb{e}_b)]^{1/2}$. After that
the extraction of $V$ from the integral for $Z_0$ leads to
$$
Z_0= N' \int D\phi Dx^\m \biggl[ {\det '(P^+_1 P_1) \over
    \det(\mb{e}_a,\mb{e}_b)} \biggr]^{1/2} e^{-I[x,g]} \, .\eqno(9)
$$
The integration over $g_{\a\b}$ is now performed only over the Weyl
slice containg ${\hat g}_{\a\b}$.

The result (9) is the first important step in the quantization
procedure, corresponding to {\it fixing the reparametrization
invariance} of the functional integral.

\subsub{The standard Faddeev-Popov.} It is illuminating to derive the
expression (8) for $Z_0$ by using the standard Faddeev-Popov approach.
We want to fix the reparametrization symmetry in the functional
integral (3b) by using the gauge condition
$$
g_{\a\b} = e^\phi {\hat g}_{\a\b}\equiv g^w_{\a\b} \, .
$$
To this end we define the Faddeev-Popov determinant by
$$
\int D\phi D\e{'}\d (g_{\a\b}-e^\phi{\hat g}_{\a\b})\D_{FP}=1 \, .
$$
Inserting this into Eq.(3b) for $Z_0$ yields
$$
Z_0= N\int D\phi D\e{'}Dg_{\a\b} Dx^\m\D_{FP}\d (g_{\a\b}
     -e^\phi{\hat g}_{\a\b}) e^{-I[x,g]} \,.
$$
The integration over $Dg_{\a\b}$ is now easily performed, leading to
the replacement $g_{\a\b}\to e^\phi{\hat g}$  in the classical action.
Writting the general gauge transformation of $g_{\a\b}$ as
$$\eqalign{
&\d g_{\a\b} =(P_1\e{'})_{\a\b} +\d\l g_{\a\b} \, , \cr
&\d\l\equiv \d\phi +\nabla\cdot\e ' \, , }
$$
the Faddeev-Popov determinant is defined by
$$
\D_{FP} = {\partial (\d\l ,P_1\e{'})\over \partial(\d\phi ,\e{'})}
      =\det \pmatrix{1 & *   \cr
                     0 & P_1 \cr} =\det{'}P_1
                                  =[\det{'}(P^+_1P_1)]^{1/2} \, ,
$$
which leads directly to Eq.(8).

\subsub{The measure $Dx^\m$.} Now we return to Eq.(9) and do the
$x^\m$ integration. The measure $Dx^\m$ is defined by the metric in
the tangent space spanned by the vectors $\d\mb{x}=(\d x^\m)$,
$$
\Vert \d\mb{x}\Vert^2 =\int d^2\xi\sqrt{g}\d x^\m\d x^\n G_{\m\n}\, ,
$$
which is invariant under reparametrizations but not under Weyl
rescalings. When a tangent vector $\d\mb{x}_0$ is $\xi$-independent
we have
$$
\Vert\d\mb{x}_0\Vert^2=\d x^\m_0\d x^\m_0 G_{\m\n}
                                              \int d^2\xi\sqrt{g}\, ,
$$
so that
$$
Dx^\m_0 =\biggl(\int d^2\xi\sqrt{g}\biggr)^{D/2}dx^1_0\dots dx^D_0\, .
$$
By using the usual decomposition
$$
x^\m =x^\m_0 + x^{\prime\m} \, ,
$$
the integration over $x^\m$ in (9) leads to
$$\eqalign{
\int Dx^\m e^{-I[x,g_w]}&=\int Dx^\m_0\int Dx^{\prime\m} \exp{
  \biggl( -\int d^2\xi\sqrt{g_w}{g}_w^{\a\b}\partial_\a
  x^{\prime\m}\partial_\b x^{\prime\n}G_{\m\n}\biggr)   } \cr
&=\int Dx^\m_0\int Dx^{\prime\m}\exp{  \biggl( \int d^2\xi\sqrt{g_w}
  x^{\prime\m}\D_{g_w} x^{\prime\n} G_{\m\n}\biggr)   }\cr
&=V_D\biggl(\int d^2\xi\sqrt{g_w}\biggr)^{D/2}
  \bigl[\det{'}(\D_{g_w})\bigr]^{-D/2}  \, .}              \eqno(10)
$$
Here, $V_D=\int dx^1_0\dots dx^D_0$ is the volume of the spacetime
$M_D$ and $\D_{g_w}$ is the covariant
d'Alambertian (the Laplace-Beltrami operator),
$$
\D_{g_w}\equiv -{1\over\sqrt{g_w}}\partial_\a\sqrt{g_w}g_w^{\a\b}
                                   \partial_\b \equiv -{\boxx}_w \, .
$$
Putting all together one easily finds
$$
Z_0= N\int D\phi \biggl[
     {\det{'}(P_1^+P_1)\over\det (\mb{e}_a ,\mb{e}_b)}\biggr]^{1/2}
     \biggl[ { \det{'}(\D_{g_w})\over \int d^2\xi\sqrt{g_w} }
                                        \biggr]^{-D/2} \, .\eqno(11)
$$
The operator $\D_{g_w}$ is not Weyl invariant, and $\det{'}(\D_{g_w})$
depends on the Weyl factor $\phi$. Thus, although the classical action
is Weil invariant, the functional integral is not, which leads to the
appearance of the conformal anomaly .

\subsub{Conformal anomaly.} In the previous equation $g^w_{\a\b}$ is the
metric on the Weyl slice. If we choose the conformally flat slice, one
can prove the relation
$$\eqalign{
&\ln \biggl[ {\det{'}(\D_{g_w})\over \int d^2\xi\sqrt{g_w} }\biggr]
 -\ln \biggl[ {\det '(\D_{\hat g})\over\int d^2\xi\sqrt{\hat g} }\biggr]
 =-{1\over 24\pi}\G [\phi ] \cr
&\G [\phi ]= \int d^2\xi \bigl[ {\fr 1 2}\d^{\a\b}\partial_\a\phi
          \partial_\b\phi +\m^2\bigl( e^\phi -1\bigr)\bigr] \, ,}
                                                           \eqno(12a)
$$
so that
$$
\biggl[ {\det{'}(\D_{g_w})\over \int d^2\xi\sqrt{g_w} }\biggr]^{-D/2}
                  =e^{{D/2\over 24\pi}\G [\phi ] } \, .    \eqno(12b)
$$

The calculation of the Faddeev-Popov determinant yields the result
of the same form with $D\to -26$, leading to the final result
$$
Z_0 = N\int D\phi \exp{\bigl[-\k \G [\phi ]\bigr]} \, ,\qquad
      \k\equiv {26-D\over 48\pi} \, .                      \eqno(13)
$$
The expression $\G [\phi ]$ characterizes the conformal anomaly of the
theory. It yields a measure of the violation of the conformal symmetry
at the quantum level. The action $\G (\phi )$ defines the {\it
Liouville theory\/}.

We are now going to make several {\it comments.}

1. In the process of calculating the final result (13) we had to
perform some renormalizations. They can be done in the standard manner
if the original Polyakov action is modified by adding two {\it
counterterms},
$$
I[x^\m ,g_{\a\b}]=\int d^2\xi\sqrt{g}g^{\a\b}\partial_\a x^\m
  \partial_\b x^\n G_{\m\n} +\m_0^2\int d^2\xi\sqrt{g}
  +\l_0\int d^2\xi\sqrt{g}R \, ,
$$
where $\m_0$ and $\l_0$ are the bare, cutoff-dependent coupling
constants. By the global Gauss-Bonnet theorem the last term is a
topological invariant and can be pulled out of the functional integral
as a factor $\exp [4\pi\l_0(2-\g )]$, showing that a sum over $\g$ is a
perturbative expansion.

2. The calculation of the anomaly can be carried out for higher genus
surfeses in a {\it similar manner\/.} In that case two metrics lying in
the space of all admissible metrics ${\cal M}_\g (\mb{g})$ are not
always connected by local diffeomorphisms and Weyl rescalings. There
exist additional degrees of freedom which label the conformal classes
of metrics, called the Teichm\"uller parameters. The functional
integral will include the summation over {\it conformally inequivalent}
classes of metrics, i.e.the integration over the Teichm\"uller
parameters. However, the anomaly is a {\it local} object, which means
that its form does not depend on the genus. For higher genus it is
convenient to use a Weyl slice which is not conformally flat. This
slice contains a reference metric $\hat{\mb{g}}$, which is chosen so as
to make the additional integration simple. There is no need to know the
explicit form of $\hat{\mb{g}}$ for the calculation of the anomaly. The
effective theory has the same form as in Eq.(13), with
$$
\G [\phi ]= \int d^2\xi \bigl[ {\fr 1 2}\hat{g}^{\a\b}\partial_\a\phi
    \partial_\b\phi + \hat{R}\phi + \m^2\bigl( e^\phi -1\bigr)\bigr]\,.
                                                           \eqno(14)
$$

3. For the critical dimension $D=26$ the $\phi$-dependence dissapears
from the functional integral, and the volume of the Weyl slice
$V_w=\int D\phi$ can be absorbed into the normalization. If $D\ne 26$
there is no Weyl symmetry at the quantum level (there is an anomaly),
and the quantization of the theory demands a special care, as discussed
by Bogojevi\'c and Sazdovi\' c at this meeting.

\subsection{2. Residual $SL(2,R)$ symmetries of the Liouville theory}
By studying the Liouville theory in the light--cone gauge Polyakov
descovered a resi\-dual $SL(2,R)$ invariance [4-6], which is very
important for understanding the structure of the theory. We shall now
discuss certain aspects of this symmetry at the Lagrangean level [7-9].

\subsub{Conformal anomaly and the Liouville action.} 
Classical action for the bosonic string is invariant under 2-d
reparametrizations and local Weyl rescalings. Quantization leads to the
appearance of an anomaly, which means that not all classical symmetries
are the symmetries of the quantum theory. We can use the
reparametrization invariance and fix the metric to the conformally flat
form
$$
g_{\a\b}(x)=e^\phi\eta_{\a\b} \,
$$
(we are now working in the Minkowski space). After that, the
integration over all metrics in the functional integral reduces to the
integration over the conformally flat gauge slice. Then, after
introducing the corresponding ghosts and using a convenient
regularization, one finds that the {\it anomaly} has the form
$$
A[\phi ,C]=\k\int d^2\xi C(\xi )(\hat{\boxx}\phi +\mu^2e^\phi )\, ,
                                                           \eqno(15)
$$
where $C(\xi )$ is the Weyl ghost, and $\hat{\boxx}=
\eta^{\a\b}\partial_\a\partial_\b$ . The anomaly is related to the
notion of the {\it effective action} by the relation
$$
\d W[\phi ]=A[\phi ,\d\phi ] \, ,                          \eqno(16)
$$
which after integration yields
$$
W[\phi ,\eta ]=\k\int d^2\xi \bigl({\fr 1 2}\phi\hat{\boxx}\phi
              +\mu^2e^\phi\bigr) \, .                      \eqno(17)
$$
Note that the quantity $\G [\phi ]$, appearing in the functional $Z_0$
in Eq.(13), is given by $\G[\phi ]=W[\phi ]-W[0]$.

The above expression for the effective action  can be easily
transformed into a covariant - looking form. By using the relations
$\sqrt{-g}R=-\hat{\boxx}\phi$ and $\sqrt{-g}{\boxx} =\hat{\boxx}$ valid
in the conformally flat gauge, one finds
$$
W[g]=\k\int d^2\xi \sqrt{-g}\biggl( {\fr 1 2}R\, {1\over{\boxx}}\, R
      +\mu^2\biggr)\, ,                                    \eqno(18a)
$$
which is a nonlocal expression, suitable for discussing the effective
action on other gauge slices. The effective action  can be written as a
local functional by introducing an auxiliary field $F$:
$$
W[F,g]=\k\int d^2\xi\sqrt{-g}\bigl( -{\fr 1 2}F{\boxx}F + FR
                                          +\mu^2\bigr)\, . \eqno(18b)
$$
The elimination of $F$ with the help of its equation of motion
leads back to (18a).

The Liouville theory is, up to a sign, defined by the effective
action $W$,
$$
I_L(F ,{g})= -\int d^2\xi\sqrt{-g}\bigl[
 {\fr 1 2}g^{\a\b}\partial_\a F\partial_\b F
 +{\a\over 2}FR + \mu^2 \bigr] \, ,                        \eqno(19)
$$
where the constant $\a$ is introduced for convenience (classically
$\a=2$), and we used $\k=1$.

\subsub{The light-cone gauge.} 
The Liouville theory simplifies significantly in the light-cone gauge,
defined by
$$\eqalign{
&g_{+-}=1 \, ,\qquad g_{--}= 0 \, , \cr
&ds^2=h(d\xi^+)^2+2d\xi^+d\xi^-\, , }                      \eqno(20)
$$
where the components of the metric tensor are given in the
light-cone coordinates $\xi^\pm=(\xi^0\pm \xi^1)/\sqrt{2}$ as
$$\eqalign{
&g_{+-}\equiv {\fr 1 2}(g_{00}-g_{11})\, ,\qquad
g_{--}\equiv {\fr 1 2}(g_{00}+g_{11})-g_{01} \, ,\cr
&h\equiv g_{++}={\fr 1 2}(g_{00}+g_{11})+g_{01} \, .}
$$

{}From the point of view of the functional integral, the use of this
gauge means that we are integrating over the light-cone gauge slice.
We expect that the amplitudes, defined by the functional integral, do
not depend on the choice of the gauge slice (the gauge independence of
the S-matrix!). Although the transition from the Weyl to the light-cone
gauge slice is admissible , it would be very interesting to check this
change of gauge in more details, as it is one of the basic consistency
requirements on the theory.

After finding the inverse metric
$$
g^{++}=0\, ,\qquad g^{+-}=g^{-+}=1\, ,\qquad  g^{--}=-h \, ,
$$
the calculation of the metric connection $\G^\a_{\b\g}$
yields the following nonvanishing components:
$$ \eqalign{
&\G_{++}^+=-{\fr 1 2}\partial_-h \, ,\qquad \G_{+-}^-={\fr 1 2}
                                                 \partial_-h \, ,\cr
&\G_{++}^-={\fr 1 2}(\partial_+h + h\partial_-h)\, ,}
$$
where $\partial_\pm =(\partial_0\pm\partial_1)/\sqrt{2}$.
The curvature components and the laplacian are of the simple form
$$\eqalign{
&R_{++}={\fr 1 2}h\partial_-^2h\, ,\qquad
      R_{+-}={\fr 1 2}\partial_-^2h\, , \qquad R=\partial_-^2h\, ,\cr
&{\boxx}=\partial_-(2\partial_+-h\partial_-)\, . }
$$

In the light-cone gauge the Liouville action becomes
$$
{\hat I}_L=\int d^2\xi\bigl[-\partial_+F\partial_-F +{\fr 1 2}
  h(\partial_-F)^2-{\fr 1 2}\a F\partial_-^2h -\mu^2\bigr] \, .
                                                           \eqno(21)
$$

The equation of motion for $F$  takes the form
$$
{\boxx} F = (\a /2)R\, ,\qquad\Rightarrow\qquad
\partial_-(2\partial_+-h\partial_-)F = (\a /2)\partial_-^2h \, .
                                                           \eqno(22)
$$
The equations of motion for $g_{\a\b}$ can be obtained from the
energy-momentum (EM) tensor. By using the relation
$$
\d \int d^2\xi\sqrt{-g}FR=-\int d^2\xi\sqrt{-g}\d g^{\a\b}
                      (\nabla_\a\nabla_\b -g_{\a\b}\nabla^2 )F \, ,
$$
one easily finds
$$
T_{\a\b}=-{\fr 1 2}\partial_\a F\partial_\b F
  +{\a\over 2}\bigl(\nabla_\a\nabla_\b -g_{\a\b}\nabla^2 \bigr) F
  +{\fr 1 2}g_{\a\b} \bigl({\fr 1 2}g^{\g\d}\partial_\g F\partial_\d F
  +\m^2 \bigr) \, .
$$
The traceless part of $T_{\a\b}$,
${\tilde T}_{\a\b} \equiv T_{\a\b}-{\fr 1 2}g_{\a\b}T$, takes the form
$$\eqalign{
&{\tilde T}_{--} = -{\fr 1 2}(\partial_-F)^2 +{\a\over 2}\partial_-^2F
                 = T_{--}  \, ,\cr
&{\tilde T}_{+-} = {\fr 1 2}hT_{--}  \, ,\cr
&{\tilde T}_{++} = {\fr 1 4}h^2{T}_{--} + (\a^2/4){\fr 1 8}
  \bigl[ (\partial_-h)^2 -2h\partial_-^2h+4\partial_-\partial_+h\bigr]
                                           \, .}           \eqno(23)
$$
Here, we used $T=-{1\over 2}(\a^2\partial_-^2h-2\m^2)$, and eliminated
$F$ by using its equation of motion. The relation
$$
\nabla_+T_{--} + h\nabla_-T_{--}= {\fr 1 2}(\a /2)^2\partial_-^3h \,
$$
leads to the important result
$$
\partial_-^3h=0\, ,                                        \eqno(24)
$$

\subsub{The Polyakov formulation.} To simplify the forhcomming
discussion we shall work with $\a /2=1$. The equation of motion for $F$
can be solved by looking at the simpler equation
$$
(2\partial_+-h\partial_-)F= \partial_-h \, .               \eqno(25)
$$
If $f$ is a solution of the homogenious equation,
$$
(2\partial_+ -h\partial_-)f=0 \, ,                         \eqno(26a)
$$
then one can check that
$$
F_0=\ln\partial_-f                                         \eqno(26b)
$$
is a particular solution of the complete equation (25), and therefore a
solution of (22).

The action (21) can be written in the form
$$
{\hat I}_L=-\int d^2\xi \bigl( {\fr 1 2}\partial_-^2hF_0
                                     +\mu^2\bigr)\, .      \eqno(27a)
$$
After going over to the $f$ variable, we find
$$ \eqalign{
{\hat I}_L[f]&=-\int d^2\xi\biggl[ -{\fr 1 2}\partial_-\bigl(
{2\partial_+f\over\partial_-f}\bigr)\partial_-\ln\partial_-f
                                                    +\mu^2\biggr]\cr
&=\int d^2\xi\biggl[ {(\partial_-\partial_+f)(\partial_-^2f)
\over (\partial_-f)^2} -{(\partial_-^2f)^2(\partial_+f)\over
(\partial_-f)^3}  -\m^2\biggr] \, .}                       \eqno(27b)
$$

It is interesting to note that by varying this action over $f$ and
after partial integrations ($\d f=0$ at the boundary) one obtains the
relation
$$
-2\partial_+\biggl( {\partial_-^2F_0\over\partial_-f}\biggr)
+\partial_-\biggl( {h\partial_-^2f_0-\partial_-^2h\over\partial_-f}
                                                    \biggr) =0 \, ,
$$
which is easily transformed into the simple form (24).

\subsub{SL(2,R) symmetry.} 
An interesting feature of the gauge-fixed Liouville action is the
presence of a residual symmetry. Let us show that the transformation
$$
\d_0f=\ve^-\partial_-f                                     \eqno(28a)
$$
is an invariance of the action (21) if the parameter $\ve^-$ satisfies
certain conditions. The above transformation implies
$$\eqalign{
&\d_0h = 2\partial_+\ve^- - h\partial_-\ve^- +\ve^-\partial_-h\, ,\cr
&\d_0F = \ve^-\partial_-F+{\a\over 2}\partial_-\ve^- \, , } \eqno(28b)
$$
Then, a direct calculation yields
$$
\d {\hat I}_L = \int d^2\xi h\partial_-^3\ve^- \, .
$$
Therefore, the transformation (28) is an invariance of the action
provided $\ve^- (\xi^+,\xi^-)$ sa\-tisfies the condition
$$
\partial_-^3\ve^- = 0 \, .                                   \eqno(29)
$$

An interesting interpretation of the nature of this
symmetry was given by Dass and Summitra [8]. The classical string
action is invariant under local reparametrizations and Weyl
rescalings:
$$\eqalign{
&\d_0g_{--}=2\partial_-\ve^-g_{--} +2\partial_-\ve^+g_{+-}
                  +\ve\cdot\partial g_{--} +\l g_{--} \, ,\cr
&\d_0g_{+-}=\partial_+\ve^+ g_{+-} + \partial_+\ve^-g_{--}
      +\partial_-\ve^+ g_{++} +\partial_-\ve^-g_{-+}
                    +\ve\cdot\partial g_{+-} +\l g_{+-}\, ,\cr
&\d_0g_{++}=2\partial_+\ve^-g_{-+} +2\partial_+\ve^+g_{++}
            +\ve\cdot\partial g_{++} +\l g_{++} \, .}
$$
Now, demanding that the above transformations do not change the
light-cone gauge, we obtain the following conditions on
the parameters $\ve^+,\ve^-$ and $\l$:
$$\eqalign{
&\d_0g_{--}=2\partial_-\ve^+ =0\, ,\cr
&\d_0g_{+-}=\partial_+\ve^+ +\partial_-\ve^- +\l =0 \, .}
$$
These conditions yield
$$
\ve^+=\ve^+(\xi^+)\, ,\qquad \l=-\partial_+\ve^+ -\partial_-\ve^- \, ,
$$
whereafter the transformation low of $g_{++}=h$ becomes
$$
\d_0h=(2\partial_+\ve^- +\ve^-\partial_-h -h\partial_-\ve^-)
       + (2\partial_+\ve^+h +\ve^+\partial_+h -h\partial_+\ve^+) \, .
$$
Finally, choosing $\ve^+=0$, we obtain  the correct form of the
symmetry transformation for $h$. The symmetry (28) is thus seen to be
{\it a remnant of the original classical string symmetry in the
light-cone gauge.}

There is an apparent contradiction between the determination of  $h$ as
the $(++)-$com\-po\-nent of the metric, and the fact that (28) is not
the tensorial transformation law. This can be easily understood if we
note that in the light--cone gauge $\det g_{\a\b}=-1$, so that $h$ can
be equally well considered as a {\it tensor density\/}. Indeed, the
transformation law (28) is in agreement with this assumption.
\vskip.2cm
{\it Exercise}. Derive the transformation law for
$G_{++}\equiv (\sqrt{g}\, g_{++})$ under reparametrizations with
$\ve^-$.
\vskip.2cm
Let us now show that the residual symmetry is $SL(2,R)$.
It is clear from Eq.(29) that $\ve^-$ must be a polinomial in $\xi^-$:
$$
\ve^-(\xi^+,\xi^-)=\o_-(\xi^+) +\xi^-\o_0(\xi^+)
                              +(\xi^-)^2\o_+(\xi^+)\, .    \eqno(30)
$$
Using this form of $\ve^-$  the transformation law  for $h$ can be
rewritten in the form
$$\eqalign{
&\d_0h=\bigl( \o_-l^- +\o_0l^0 +\o_+l^+\bigr)h +2\bigl(
\partial_+\o_- +\xi^-\partial_+\o_0 +(\xi^-)^2\partial_+\o_
                                                   +\bigr) \, ,\cr
&l^-\equiv \partial_- \, ,\qquad l^0\equiv \xi^-\partial_- -1 \, ,
       \qquad l^+\equiv (\xi^-)^2\partial_--2\xi^- \, .}   \eqno(31)
$$
The quantities $l^a$ are the generators of the $SL(2,R)$ algebra (see
Appendix C):
$$
\bigl[l^0,l^-\bigr]=-l^-\, ,\qquad\bigl[l^0,l^+\bigr]=l^+\, ,\qquad
\bigl[l^+,l^-\bigr]=-2l^0\, .                              \eqno(32)
$$
Thus, up to the $\partial_+\o_a$ terms, the transformation of $h$ is an
$SL(2,R)$ transformtion.

A more transparent way of seeing the nature of the transformation law
of $h$ is the following. The solution of the equation of motion (24)
for $h$ is a polinomial in $\xi^-$:
$$
h(\xi^+,\xi^-)=J^+(\xi^+) -2\xi^-J^0(\xi^+) +(\xi^-)^2J^-(\xi^-)\, .
                                                           \eqno(33)
$$
The transformation low for $J^a$ is given by
$$\eqalign{
&\d_0J^+=-2\o_-J^0 -\o_0J^+ +2\partial_+\o_- \, ,\cr
&\d_0J^0=\o_+J^+ -\o_-J^- -\partial_+\o_0 \, ,\cr
&\d_0J^-=\o_0J^- +2\o_+J^0 +2\partial_+\o_+ \, .}          \eqno(34a)
$$
By using the results of Appendix C, one can rewrite the above
relations in the form
$$
\d_0J^a =  f^{abc}\o_bg_{cd}J^d +2g^{ab}\partial_+\o_b \, ,\eqno(34b)
$$
where the $SL(2,R)$ character of $J^a$ and $\o_b$ is explicit.

It is interesting to note that after using $T_{--}=0$, the traceless
EM tensor $\tilde T_{++}$ can be written in terms of the
quantities $J^a$ as follows:
$$
\tilde T_{++}^y ={\fr 1 4}[2(J^0)^2 - J^+J^- - J^-J^+]
                 +(-\partial_+J^0 + \xi^-\partial_+J^-)\, .\eqno(35a)
$$
Now, the equation  $T=0$ leads to $J^-=\m^2$, and, consequently,
to $\partial_+J^-=0$, so that the EM tensor can be written in the
Sugawara form:
$$
\tilde T_{++}^y = {\fr 1 4}g_{ab}J^aJ^b -\partial_+J^0 \, .\eqno(35b)
$$

\subsection{3. Equivalence between conformal and light-cone gauge} 
We shall now consider a relation beteen the conformal and the
light-cone gauge, and see how one can derive Polyakov's action and the
EM tensor in the light-cone gauge from the corresponding expressions in
the conformal gauge [10-12].

\subsub{1.} In the conformal gauge the metric is defined by
$$
ds^2=2e^{\phi (x)}dx^+ dx^- \, ,
$$
whereas in the light-cone gauge it takes the form
$$
ds^2=h(y)dy^+ dy^+ + 2dy^+ dy^- \, .
$$
It is easy to see that the coordinate transformation
$$
x^+=y^+ \, ,\qquad x^- = f(y^+,y^-) \, ,                   \eqno(36a)
$$
transforms the first metric into the second one provided the function
$f(y)$ satisfies the conditions
$$
h(y)=2 e^{\phi (x)}\partial_+f(y) \, ,\qquad
1=  e^{\phi (x)}\partial_-f(y) \, .                        \eqno(36b)
$$
{}From here we obtain the relation
$$
h=2{\partial_+f\over \partial_-f} \, ,                     \eqno(36c)
$$
that has been extensively used by Polyakov in his study of the
light-cone gauge dynamics. We see that the function $f(y)$ has a clear
geometrical meaning. The above coordinate transformation will be used
to relate certain results on the induced 2-d gravity in the conformal
and the light-cone gauge, respectively.

\subsub{2.} The local form of the Liouville action is given by
Eq.(19). To simplify the forthcomming expressions we shall work with
$\a /2=1$. By using the equation of motion for $F$ and going over to
the conformal gauge,
$$
\boxx F=R\qquad\Longrightarrow\qquad \hat{\boxx}F=-\hat{\boxx}\phi\, ,
$$
the Liouville action takes the form
$$
I_L = -\int d^2 x\biggl( -{\partial F\over\partial x^+}
      {\partial F\over\partial x^-}+\m^2 e^\phi \biggr)\, .\eqno(37)
$$
In what follows, we shall use the notation $F\equiv F(x)$, $F'\equiv
F(y)$, and assume that $F$ is a scalar function, i.e. $F'=F$.  By using
the transformation rule (36) the action (37) goes over into the
expression
$$\eqalign{
I_L&=-\int d^2y (\partial_-f)\bigl[ -{\fr 1 2}\bigl( 2\partial_+
            -h\partial_- \bigr) F'(\partial_-f)^{-1}\partial_- F'
                                           +\m^2 e^{\phi}\bigr] \cr
   &=-\int d^2y\bigl[ -{\fr 1 2}\partial_-h\partial_-F_0
                                     +\m^2\bigr] \, ,}     \eqno(38)
$$
which is the Liouville action in the light-cone gauge, as demonstrated
in sect. 2.

\subsub{3.} The EM tensor corresponding to the Liouville action in the
conformal gauge takes the form
$$\eqalign{
&T_{++}^x=-{\fr 1 2}\biggl( {\partial F\over\partial x^+} \biggr) ^2
          +\nabla_+^2 F \, ,\cr
&T_{--}^x=-{\fr 1 2}\biggl( {\partial F\over\partial x^-} \biggr) ^2
          +\nabla_-^2 F \, ,\cr
&T_{+-}^x=-\nabla_+\nabla_- F +{\fr 1 2}\m^2e^\phi \, , }  \eqno(39a)
$$
where
$$
\nabla_\pm^2F=(\partial_\pm -\partial_\pm\phi )\partial_\pm F\, ,
\qquad \nabla_+\nabla_- F=\partial_+\partial_- F\, ,
$$
and $\partial_\pm F$ can be replaced by $-\partial_\pm\phi$.
The equation $T_{+-}^x=0$ is the well known Liouville equation.

The transition to the light-cone gauge is realised with the help of
the coordinate transformation (36),
$$
T_{\a\b}(y)={\partial x^\g\over\partial y^\a}
            {\partial x^\d\over\partial y^\b} T_{\g\d}(x) \, ,
$$
which leads to
$$\eqalign{
&T_{++}^y = T_{++}^x +h(\partial_-f) T_{+-}^x
           +{\fr 1 4}h^2(\partial_-f)^2 T_{--}^x \, ,\cr
&T_{--}^y = (\partial_-f)^2 T_{--}^x \, ,\cr
&T_{+-}^y = (\partial_-f) T_{+-}^x
             +{\fr 1 2}h(\partial_-f)^2 T_{--}^x \, .}     \eqno(40)
$$
The components $T_{\a\b}(x)$, when expressed in the light-cone gauge
coordinates $y$, take the form
$$\eqalign{
&T_{++}^x = {\fr 1 8}\bigl[ (\partial_-h)^2 -2h\partial_-^2h\bigr]
           -{\fr 1 2}\partial_+\partial_-h \, ,\cr
&(\partial_-f)^2T_{--}^x =-\partial_-^2\ln\partial_-f
    +{\fr 1 2}(\partial_-\ln\partial_-f)^2 \equiv -\{ f,y^-\}\, ,\cr
&(\partial_-f)T_{+-}^x = -{\fr 1 2}\partial_-^2h
                                       +{\fr 1 2}\m^2 \, ,} \eqno(41)
$$
where $\{ f,y^-\}$ denotes the Schwarz derivative.

Introducing the traceless tensor
$\tilde T_{\a\b}^y $, the above equation can be rewritten as
$$\eqalign{
&\tilde T_{++}^y = T_{++}^y -{\fr 1 2}hT^y =
                   T_{++}^x +{\fr 1 4}h^2 T_{--}^y\, ,\cr
&\tilde T_{--}^y =  T_{--}^y \, ,\qquad \tilde T_{+-}^y =  0\, .}
                                                           \eqno(42)
$$

By using the solution of the equation of motion $\{f,y^-\}=0$,
$$
f(y^+,y^-)={a(y^+)y^- +b(y^+)\over c(y^+)y^- +d(y^+)} \, , \qquad
       ad-bc=1 \, ,                                        \eqno(43)
$$
(which is $SL(2,R)$ invariant) the corresponding expression for $h(y)$
takes the form (33), where the currents $J^a$ are determined in terms
of the coefficients $a,b,c$ and $d$.
Then, a direct calculation of $\tilde T_{++}^y$, with the help of the
equation $T_{+-}^y=0$, leads to the effective EM tensor as in Eq.(35a).
Thus, the solution (43) correctly reveals the $SL(2,R)$ structure of
the theory.

\subsection{4. Concluding remarks} 

We have seen that $SL(2,R)$ is the residual symmetry of the gauge-fixed
Liouville action. However,  $SL(2,R)$ is not the symmetry of the
complete Liouville theory in the light-cone gauge [8,9], as the
gauge-fixed action is not equivalent to the gauge-fixed theory.  The
essential features of the residual $SL(2,R)$ symmetry can be
successfully treated in the Hamiltonian framework [13-14]. This is an
important step in understanding the full quantum Liouville theory
[15,16].

\subsection{Appendix} 

\subsub{A. Geometry of surfaces}. Here we shall state several
mathematical theorems (non-rigorously and without the proof) which will
be very usefull in discussing the functional integral $Z_0$ in Section
1.

{\bf T1.} Any compact, oriented, closed and connected two-dimensional
manifold without boundaries is {\it topologically equivalent} to a
sphere with $\g$ handles. The genus $\g$ is related to the Euler
characteristic $\chi$ by the relation (the global Gauss-Bonnet theorem)
$$
{1\over 4\pi}\int d^2\xi \sqrt{g} R \equiv \chi =2-2\g  \, .
$$

{\bf T2.} A manifold $\S$ endowed with a metric $\mb{g}=(g_{\a\b})$ is
called the {\it metric space}. Any metric on $\S$ can be {\it locally}
put into the conformally flat form
$$
g_{\a\b}(\xi )=e^{\phi (\xi )}\d_{\a\b} \, ,
$$
by a suitable coordinate transformation.

By a Weyl rescaling we can, again locally, reduce the metric to the
Euclidean form, $g_{\a\b}=\d_{\a\b}$. For $\g\ne 1$ this cannot be done
globally; indeed, globaly flat metric implies $R=0$ and, therefore,
$\chi =0$, which is possible only if $\g =1$.

{\bf T3.} Any oriented two-dimensional metric space $(\S,\mb{g})$ is a
complex manifold.

{\bf T4.} Let $\S$ be a compact, oriented and closed two-dimensional
manifold of genus $\g$, with a metric \mb{g}. Then, there exists a
(nonsingular) combination of diffeomorphisms (local $+$ global !) and
Weyl rescalings that transforms the metric \mb{g} into one of the
following canonical forms (characterized by the constant curvature),
$$\eqalign{
&R=+1\, ,\qquad ds^2={\vert dz\vert^2 \over (1+\vert z\vert^2)^2 } \, ,
                                        \,\, \qquad \g =0 \, ,\cr
&R=0\, ,\phantom{+}\qquad ds^2 =\vert dz\vert^2 \, ,
                                   \hskip1.8cm \g =1 \, , \cr
&R=-1\, ,\qquad ds^2={\vert dz\vert^2\over \vert Im\, z\vert^2 }\, ,
                                  \,\,\hskip1.3cm \g \ge 2\, ,}
$$
whereas the manifold $\S$ goes over into
$$\eqalign{
&\hat C \, ,  \qquad  \qquad  \g =0 \, , \cr
&C/\G_1 \, ,  \;\,    \qquad  \g =1 \, ,\cr
&U/\G_\g \, , \;\,    \qquad   \g\ge 2 \, . }
$$
Here, $\hat C$ is the compactified complex plane (the Riemann sphere),
$C$ is the complex plane, $U$ is the open upper half plane excluding
$z=\infty \,$, while $\G_\g$ are some discrete groups of conformal
transformations, which leave the canonical metrics invariant. The
notation $C/\G_1$ means that the points in $C$ which are connected by
$\G_1$ should be identified, and similarly for $U/\G_\g$.

One can show that there exists the fundamental region in $C$ $(U)$ such
that any point in $C$ $(U)$ can be obtained from this fundamental
region by applying an element of $\G_\g$.  Note that $\G_0$ is trivial,
i.e. the manifold $\S_0$ can be obtained from the Riemann sphere $\hat
C$ by using {\it only Weyl rescalings and local diffeomorphisms}.

{\it Comment}. The group of all diffeomorphisms $Diff$ cansists of the
local diffeomorphisms $Diff_0$ (which are conected to the identity) and
the global ones (which are not).

{\bf T5.} The number of the conformal Killing vectors $N'_\g $ of the
metrics (6a) is given by the relation
$$\eqalign{
&N'_0  =6 \, , \qquad \qquad  \g =0 \, , \cr
&N'_1  =2 \, , \qquad \qquad  \g =1 \, , \cr
&N'_\g =0 \, , \qquad \qquad  \g\ge 2 \, . }
$$
The group $\G_\g$ is uniquely determined by a finite number $N_\g$
of real Teichm\"uller parameters. For all $\g$ we have a relation
$$
 N_\g - N'_\g = 6\g -6 \, ,
$$
which is known as the {\it Riemann-Roch theorem}.

\subsub{B. General form of the effective action} 
The effective action (3) is calculated with respect to the flat
background metric $\hat g_{\a\b}=\eta_{\a\b}$. We shall find out
here its form with respect to an arbitrary background $\hat g_{\a\b}$.

First we note that the anomaly can be written in the covariant form
$$
A=\k\int d^2\xi C(\xi )\sqrt{-g}\bigl(-R(g)+\mu^2\bigr) \, .
$$
This follows from Eq.(1) by noting that $\sqrt{-g}R(g)=
-\hat{\boxx}\phi$ and $\sqrt{-g}=e^\phi$ in the conformally flat gauge.
Now we can generalize the gauge condition to the form,
$$
g_{\a\b}=e^\phi\hat g_{\a\b}\, ,
$$
(which, in the functional integral, corresponds to integrating over an
arbitrary Weyl slice in the space of all metrics) so that, after using
the identity
$$
\sqrt{-g}R=\sqrt{-\hat g}(\hat R-\hat{\boxx}\phi )\, ,
$$
the anomaly becomes
$$
A[\phi ,\hat g]=\k\int d^2\xi C\bigl( \xi )\sqrt{-\hat g}
                  (\hat{\boxx}\phi -\hat R+\mu^2e^\phi\bigr)\, .
$$
Integrating the relation (2) we obtain the effective action in the form
$$
W[\phi ,\hat g]=\k\int d^2\xi\sqrt{-\hat g}\bigl( {\fr 1 2}
      \phi\hat{\boxx} \phi -\phi\hat R+\mu^2e^\phi\bigr)\, , \eqno(B1)
$$
which is a natural generalization of Eq.(3).

\subsub{C. The $SL(2,R)$ algebra.} 
The basic matrix representation of the group $SL(2,R)$ consists of all
real matrices $M$ of order two, with $\det M=1$. The corresponding
generators are real, traceless $2\times 2$ matrices. Let us choose a
basis of generators
$$
T_2 = {\fr 1 2}\s_1 ={\fr 1 2}\pmatrix{0& 1\cr
                                       1& 0\cr} \, ,\qquad
J_0 = {\fr 1 2}i\s_2 ={\fr 1 2}\pmatrix{ 0& 1\cr
                                        -1& 0\cr} \, ,\qquad
T_1 = {\fr 1 2}\s_3 ={\fr 1 2}\pmatrix{1&  0\cr
                                       0& -1\cr} \, ,
$$
where $\s_i$ are the Pauli matrices. The commutation rules
$$
[J_0,T_2]=T_1\, ,\qquad [J_0,T_1]=-T_2\, ,\qquad [T_1,T_2]=J_0 \, ,
$$
define $sl(2,R)$, the Lie algebra of $SL(2,R)$. In the $(+-)$ basis
$$
G_\pm=J_0 \pm T_2 \, ,\qquad G_0 =2T_1\, ,
$$
the commutation relations take the form
$$
[G_0,G_+]=2G_+\, ,\qquad [G_0,G_-]=-2G_-\, ,\qquad [G_+,G_-]=-G_0\, .
                                                            \eqno(C1)
$$
It is now easy to find the structure constants of the algebra,
$$\eqalign{
&[G_a,G_b]\equiv f_{ab}{^c}G_c \, ,\cr
&f_{0+}{^+}=2\, ,\qquad f_{0-}{^-}=-2 \, ,\qquad f_{+-}{^0}=-1 \, ,}
$$
and the related Cartan's metric,
$$
g_{ab}\equiv -Tr\bigl( G_aG_b\bigr) =
           \pmatrix{0&   0&   1\cr
                    0&  -2&   0\cr
                    1&   0&   0\cr} \qquad (a,b=+,0,-)\, . \eqno(C2)
$$
Since the metric $g_{ab}$ is nonsingular, one can calculate its inverse
$g^{ab}$ and define completly antisymmetric structure constants:
$$
f^{abc}\equiv g^{ad}g^{be}f_{de}{^c}\, ,\qquad f^{+-0}=1\, .\eqno(C3)
$$

The values of the structure constants and metric depend on the
choice of basis. The quadratic invariant takes the correct form
$$
g^{ab}G_aG_b=2(G_+G_-) -{\fr 1 2}(G_0)^2  =
                  2\bigl[(J_0)^2-(T_2)^2-(T_1)^2\bigr] \, ,
$$

\subsection{References}

\item{1.}\aut{B. Hatfield}, \knj{Quantum Field Theory of Point
Particles and Strings} (Adisson-Wesly, Readwood City, CA, 1992).
\item{2.}\aut{S. Weinberg}, \rad{Covariant Path-Integral Approach to
String Theory}, Lectures at the $3^{rd}$ Jerusalem Winter School of
Theoretical Physics, University of Texas preprint UTTG-17-87 (1987).
\item{3.}\aut{E. D'Hoker and D. H. Phong}, \rad{The geometry of string
perturbation theory}, Rev. Mod. Phys. {\bf 60} (1988) 917.
\item{4.}\aut{A. M. Polyakov}, \rad{Quantum Gravity in Two Dimensions},
Mod. Phys. Lett. {\bf A2} (1987) 893.
\item{5.}\aut{V. G. Knizhnik, A. M. Polyakov and A. B. Zamolodchikov},
\rad{Fractal Structure of 2d-Quantum Gravity}, Mod. Phys. Lett. {\bf
A3} (1988) 819.
\item{6.}\aut{A. Polyakov}, in: \knj{Fields, Strings and Critical
Phenomena}, Les Houches Lectures (1988), eds. E. Brezin and
J. Zinn-Justin (Elsevier Science Publishers, 1989).
\item{7.}\aut{N. D. Hari Dass and R. Sumitra}, \rad{Symmetry
Reorganization in Exactly Solvable Two-Dimensional Quantized
Supergravity}, Int. J. Mod. Phys. {\bf } (1988) 2245.
\item{8.}\aut{J. A. Helayel-Neto, S. Mokhtari and A. W. Smith},
\rad{Quantum Gravity in Two Dimensions and SL(2,R) Current Algebra},
Phys. Lett. {\bf B236} (1990) 12.
\item{9.}\aut{Kai-Wen Xu and Chuan-Jie Zhu}, \rad{Symmetry in
two-dimensional gravity}, Int. J. Mod. Phys. {\bf A6} (1991) 2331;
\item{10.}  F. David, Mod. Phys. Lett. {\bf A3} (1988) 1651;
J. Diestler and H. Kawai, Nucl. Phys. {\bf B321} (1989) 509.
\item{11.}\aut{Chang-Jun Ahn, Young-Jai Park, Kee Yong Kim, Yongduk
Kim, Won-Tae Kim and Byung-Ha Cho}, \rad{Two-dimensional quantum
gravity in the conformal gauge}, Phys. Rev {\bf D42} (1990) 1144.
\item{12.}\aut{Chang-Jun Ahn, Won-Tae Kim, Young-Jai Park, Kee Yong
Kim and Yongduk Kim}, \rad{Equivalence of light-cone and conformal
gauges in two-dimensional quantum gra\-vity}, Mod. Phys. Lett.
{\bf A7} (1992) 2263.
\item{13.}\aut{Ed. Sh. Egorian and R. P. Manvelian}, \rad{Canonical
formulation of 2D induced gravity}, Mod. Phys. Lett. {\bf A5} (1990)
2371; \aut{E. Abdalla, M. C. B. Abdalla, J. Gamboa and A. Zadra},
\rad{Gauge-independent analysis of 2D-gravity}, Phys. Lett.
{\bf B273} (1991) 222;
\aut{J. Barcelos-Neto}, \rad{Constraints and hidden symmetry in
2D-gravity}, Univ. Rio de Janeiro preprint IF/UFRJ/92/21 (1992).
\aut{S. Ghosh and S. Mukherjee}, \rad{$SL(2,R)$ currents in
2D-gravity are generators of improper gauge transformations}, Saha
Institute preprint (1993).

\item{14.}\aut{M. Blagojevi\'c, M. Vasili\'c and T. Vukasinac},
\rad{Hamiltonian analysis of $SL(2,R)$ symmetry}, Institute of
Physics preprint IF-11/93 (1993).

\item{15.}\aut{A. Mikovi\' c}, \rad{Canonical quantization approach
to 2d gravity coupled to $c<1$ matter}, Imperial College preprint
Imperial-TP/92-93/15 (1993).
\item{16.}\aut{T. Kuramoto}, \rad{BRST Quantization of Conformal Field
Theories on a Random Surface}, Phys. Lett. {\bf B233} (1989);
\aut{Y. Tanii}, \rad{On the BRST Approach to Two-Dimensional
Gravity in the Light-Cone Gauge}, Int. J. Mod. Phys. {\bf A6} (1991)
4639.

\bye